\documentclass[aps,prl,twocolumn,superscriptaddress,longbibliography]{revtex4-2}

\usepackage{amsmath,amssymb,bm,bbm,braket,graphicx,hyperref}
\hypersetup{colorlinks=true,linkcolor=blue,citecolor=blue,urlcolor=blue}

\newcommand{\vect}[1]{\bm{#1}}

\usepackage[english]{babel}
\usepackage{threeparttable}
\usepackage{graphicx}
\usepackage{caption}
\usepackage[caption=false,position=top,singlelinecheck=off,justification=raggedright]{subfig}
\usepackage {cancel}
\usepackage{comment}
\usepackage{caption}
\usepackage{mathtools}

\usepackage{mathdots}
\usepackage[utf8]{inputenc}
\usepackage{times}
\usepackage{newtxtext}
 
\usepackage{multibib}
\newcites{appx}{References}

\usepackage{mathptmx}
\usepackage{amsthm}

\usepackage{caption}
        \usepackage[top=30mm, bottom=30mm, left=25mm, right=25mm]{geometry}
 \usepackage{amsmath}
        \usepackage{amsmath, amssymb}
	\usepackage{color}
	\usepackage{graphicx}
	\usepackage{braket}
	\usepackage[subrefformat=parens]{subcaption}
	\usepackage{txfonts}
	\usepackage{enumerate}
	\usepackage{bm}
	 
 \usepackage{ragged2e}

	\definecolor{gray}{gray}{0.5}
	\definecolor{emerald}{cmyk}{1,0,0.5,0}
	\definecolor{bluegreen}{cmyk}{0.85,0,0.33,0}
	\definecolor{violet}{cmyk}{0.79,0.88,0,0}

\usepackage{soul}
\usepackage{ulem}
\usepackage{lipsum}
\usepackage{footnote}

\usepackage{amsmath}

\begin{document}
\preprint{APS/123-QED}

\title{Quantum Spin Squeezing Enhanced by Critical Exceptional Points}

\author{Yuma Nakanishi}
\affiliation{Department of Physics, University of Tokyo, 7-3-1 Hongo, Bunkyo-ku, Tokyo 113-0033, JAPAN}

  \email{nakanishi.y@phys.s.u-tokyo.ac.jp}

\date{\today}

\begin{abstract}
Critical exceptional points (CEPs) are nonequilibrium critical points in open many-body systems at which multiple collective excitation modes coalesce. CEPs are known to amplify classical fluctuations, but their effect on genuinely \textit{quantum} fluctuations remains unclear.
Here, we show that dissipative collective-spin systems hosting CEPs exhibit parametrically enhanced steady-state \textit{quantum} spin squeezing. 
Close to the CEP, the optimally squeezed variance scales as $|Z|$, whereas the anti-squeezed variance diverges as $|Z|^{-1}$, with $Z$ the dimensionless order parameter. Importantly, the anti-squeezed fluctuation direction asymptotically aligns with the coalescing eigenvector of the stability matrix, reflecting the defective nature of the CEP dynamics.
These scalings are robust against dephasing channels generated by spin components orthogonal to the coalesced critical collective mode.
Our results identify CEPs as a route to engineering steady-state anisotropic quantum fluctuations and correlations in driven-dissipative platforms.
\end{abstract}

\maketitle

\begin{figure}[t]
  \centering
\includegraphics[width=1.0\linewidth]{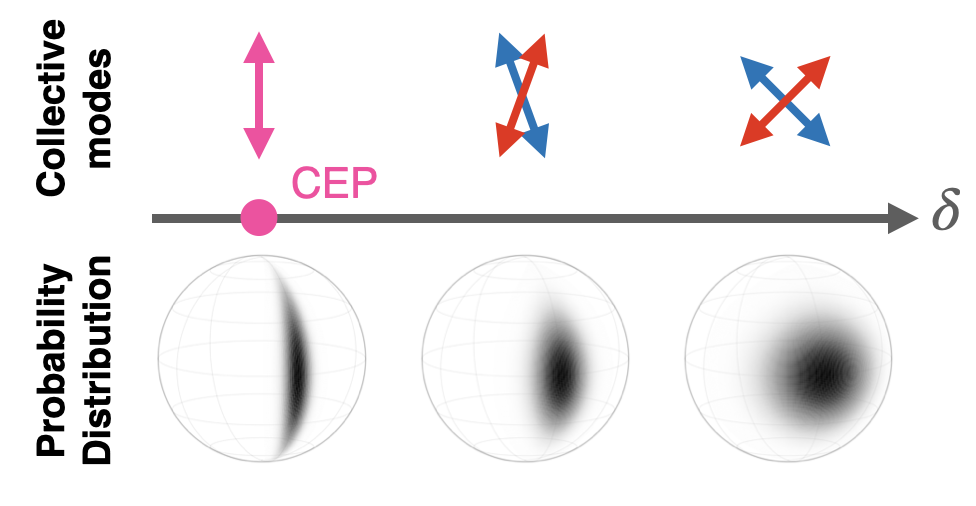}
\caption{\justifying\small 
Top: schematic of collective excitation modes for the control parameter $\delta$. The two linear collective modes (arrows) approach each other and coalesce at the CEP (pink) into a single defective direction.
Bottom: schematic illustration of the steady-state spin distribution on the Bloch sphere. Near the CEP, the distribution becomes strongly anisotropic and elongated along the defective direction, whereas away from the CEP it becomes progressively more isotropic.
}
  \label{fig:schematic}
\end{figure}
\noindent\textit{Introduction.—}
Open quantum systems exhibit \textit{nonequilibrium} phase transitions that cannot be captured within equilibrium statistical mechanics based on free-energy minimization and detailed balance~\cite{Diehl2008QuantumStates,Sieberer2013DynamicalCriticalPhenomena,Sieberer2016Keldysh,Minganti2018LiouvillianSpectralTheory}. 
In such systems, quantum coherence and quantum fluctuations do not merely survive passively on top of this nonequilibrium dynamics but often cooperate with it to determine the phase of matter and criticality themselves. 
As a result, open quantum systems provide a natural arena in which nonequilibrium many-body physics and genuinely quantum effects combine to generate a new critical behavior~\cite{DallaTorre2010QuantumCriticalNoise,Marcuzzi2016AbsorbingState,Buchhold2017NonequilibriumEFT, Marino2016DrivenMarkovianQuantumCriticality,MarinoDiehl2016QuantumDynamicalFieldTheory,RussoPohl2025PRL_QDCTC}.

In open systems, the non-Hermitian nature of the dynamical generator can underlie the emergence of exceptional points (EPs), which are non-Hermitian degeneracies at which eigenvalues and eigenmodes coalesce~\cite{Kato,Heiss,Ashida2020AdvPhys}. 
In their vicinity, a variety of nontrivial phenomena can occur, such as exotic lasing phenomena~\cite{feng2014single, hodaei2014parity}, loss-induced transparency~\cite{guo2009observation}, and enhanced sensitivity~\cite{zhang2019quantum, hodaei2017enhanced, wang2019arbitrary}.
EPs can also shape critical phenomena and phase transitions~\cite{ashida2017parity}. 
A particularly important case arises in nonequilibrium steady-state phase transitions without detailed balance, where the critical sector of the linearized macroscopic stability matrix becomes defective. 
Such points are called \textit{critical exceptional points (CEPs)}~\cite{Hanai2019PRL, Fruchart2021Nature,HanaiLittlewood2020PRR,Belyansky2025PRL,Nakanishi3,Zelle2024PRX,Suchanek2023PRL,Suchanek2023PRE,you2020nonreciprocity,Nakanishi_review}.
The resulting zero-eigenvalue Jordan block in the nonequilibrium stability matrix makes the steady state unusually susceptible to perturbations, leading to strongly anisotropic and enhanced fluctuations.
This anomalous fluctuation amplification has several important consequences: an increase in the effective upper critical dimension~\cite{HanaiLittlewood2020PRR}, divergent entropy production~\cite{Suchanek2023PRL,Suchanek2023PRE}, and fluctuation-induced first-order transitions~\cite{Zelle2024PRX}.
However, it remains unclear whether the CEP leaves a distinct imprint on \textit{quantum} fluctuations.

In this work, we address this question by studying the impact of CEPs on open quantum systems, focusing on \textit{quantum spin squeezing}. Since spin squeezing directly probes the suppression of quantum fluctuations along an optimal collective-spin direction, it provides a natural probe of CEP-induced anisotropy in quantum fluctuations. Moreover, spin squeezing is not merely a diagnostic of critical fluctuations: it is an experimentally accessible witness of entanglement and can also support quantum-enhanced measurements~\cite{Wineland1994PRA,Sorensen2001PRL,Pezze2018RMP}. 
Spin squeezing has already been demonstrated experimentally in Bose--Einstein condensates \cite{Esteve2008Nature,Gross2010Nature,Riedel2010Nature,Muessel2014PRL}, cavity-mediated and free-space atomic ensembles \cite{Leroux2010PRL,Hosten2016Nature,Sewell2012PRL}, trapped-ion crystals \cite{Bohnet2016Science}, neutral-atom arrays \cite{Bornet2023Nature,Hines2023PRL}, in optical-clock platforms \cite{Pedrozo2020Nature,Eckner2023Nature,Robinson2024NatPhys}, and ensembles of nitrogen-vacancy centers in diamond \cite{Wu2025Nature}.

We reveal that dissipative collective spin systems hosting CEPs generically exhibit \textit{steady-state} spin squeezing (Figure~\ref{fig:schematic}).
Specifically, for a \(\mathcal{PT}\)-symmetric class of collective-spin systems, we analytically show that, near the CEP and in the thermodynamic limit, the spin-squeezing parameter generically scales as \(|Z|\), whereas the corresponding anti-squeezing scales as \(|Z|^{-1}\), where \(Z\) is the dimensionless order parameter. Moreover, the anti-squeezing direction coincides with the coalescence direction and remains robust against additional decoherence generated by spin components orthogonal to it.
These results are obtained by solving the Lyapunov equation for the covariance matrix. 
In a representative example, we verify our analysis and further examine the quantum spin entanglement and finite-size scaling of the spin-squeezing parameter.
Our results identify CEPs as a promising route to controlling steady-state quantum fluctuations and correlations in driven--dissipative platforms.

\noindent\textit{Quantum spin squeezing.—}
For a system of $N$ spin-$1/2$ particles, we introduce the collective spin operators
$S_\alpha=\frac12\sum_{i=1}^N \sigma_i^\alpha$ $(\alpha=x,y,z)$
and denote $\bm S=(S_x,S_y,S_z)^\top$.
The Kitagawa-Ueda squeezing parameter is defined from the minimal variance of a spin component in the plane perpendicular to the mean spin $\langle\bm S\rangle$:
\begin{equation}
\label{KU}
\xi_S^2 \equiv \frac{2\,\min(\Delta S_\perp)^2}{S},
\end{equation}
where $(\Delta S_\perp)^2$ denotes the minimum variance of a collective-spin component over all directions perpendicular to $\langle\bm S\rangle$, and $S=N/2$ is the total spin.
A state is spin-squeezed when $\xi_S^2<1$.

\noindent\textit{GKSL equation.—} We consider the fully-connected identical $N$ spin-$1/2$ particles governed by the Gorini--Kossakowski--Sudarshan--Lindblad (GKSL) master equation~\cite{Gorini1976JMP,Lindblad1976CMP}
\begin{equation}
\dot\rho=\hat{\mathcal{L}}\ [H,\{L_\mu\}]\rho := -i[H,\rho] + \sum_\mu \Big( L_\mu \rho L_\mu^\dagger - \frac{1}{2}\{L_\mu^\dagger L_\mu,\rho\} \Big),
\label{eq:GKSL}
\end{equation}
where $H$ is the Hamiltonian and $\{L_\mu\}$ are Lindblad operators that encode dissipation channels. Here, we write the superoperator as $\hat{\mathcal{L}}\ [H,\{L_\mu\}]$ to make its dependence on $H$ and $\{L_\mu\}$ explicit. This superoperator is called the Lindbladian.

We work with the intensive collective-spin operators
$m_\alpha:=S_\alpha/S$ $(\alpha=x,y,z)$
and $\vect{m}:=\bm{S}/S$.
We restrict ourselves to an extensive Hamiltonian that is quadratic and to Lindblad operators that are linear in the collective spin,
\begin{equation}
H=S\ \Big(\vect{B}\!\cdot\!\vect{m}+\sum_{i,j}K_{ij}\,{\{m_i,m_j\}}\Big),
\quad
L_\mu=\sqrt{S}\,\vect{\ell}_\mu\!\cdot\!\vect{m},
\label{eq:quadH_linL}
\end{equation}
with real $\vect{B}$, real-symmetric $K$, and complex $\vect{\ell}_\mu$.
We define mean value $\vect{M}=(X,Y,Z):=\langle\vect{m}\rangle$ and scaled fluctuations
$\eta_i=\sqrt{N}(m_i-M_i)$ with symmetrized covariance $\Sigma_{ij}=\tfrac12\langle\{\eta_i,\eta_j\}\rangle$, $(i,j=x,y,z)$.
Upon restricting ourselves to the class of models defined in Eq.~\eqref{eq:quadH_linL}, the Gaussian description is asymptotically exact at leading order in $1/S$: the first moments $\dot{\vect{M}}={\bf{g}}(\vect{M})$ and the symmetrized covariance matrix $\Sigma$ form a closed set and obey~\cite{CarolloLesanovsky2024PRL,DuboisSaalmannRost2021,Fiorelli2024JPhysA}
\begin{equation}
\dot{\Sigma}=D+J\Sigma+\Sigma J^T,
\qquad
J_{ij}=\partial g_i/\partial M_j,
\label{eq:Lyapunov}
\end{equation}
where $J$ denotes the Jacobian matrix, while $D$ is the diffusion matrix, determined by the Lindblad operators $\{L_\mu\}$.
Let \(R\) be the rotation that aligns the mean spin \(\bm{M}\) with the \(z\) axis, and define the rotated covariance matrix as \(\Sigma' = R \Sigma R^T\).
In the rotated frame, the transverse block \(\Sigma'_{\perp}\) of the spin covariance matrix directly determines the Kitagawa--Ueda squeezing parameter, $\xi_S^2=\lambda_{\min}({\Sigma}^\prime_\perp)$, 
where \(\lambda_{\min}\) denotes the smallest eigenvalue~\cite{Kitagawa1993PRA}.

\noindent\textit{Lindbladian $\mathcal{PT}$ symmetry and critical exceptional point.—}
To discuss the emergence of CEPs in a controlled setting, we restrict ourselves to Lindbladians that are invariant under a $\mathcal{PT}$ transformation~\cite{Huber1,Huber2,Nakanishi3,Nakanishi_review,Nakanishi1,Nakanishi2},
\begin{equation}
\hat{\mathcal{L}}\ [\mathbb{PT}(H),\{\mathbb{PT}(L_\mu)\}]=\hat{\mathcal{L}}\ [H,\{L_\mu\}],
\label{HuberPT}
\end{equation}
where we define the $\mathbb{PT}$ action on an operator $O$ as
$\mathbb{PT}(O)\equiv(PT)\,O^\dagger\,(PT)^{-1}$,
with $T$ the complex-conjugation operator. For concreteness, we use \(P=\prod_i \sigma_i^x\) as a representative parity operator; the results are independent of this choice of spin basis.
This symmetry guarantees that the associated mean-field dynamics on the Bloch sphere inherits a nonlinear PT symmetry (n-PT)~\cite{Nakanishi3,Nakanishi_review}, ${\bf g}({\tilde{P}\tilde{T}\vect M})=-\tilde{P}\tilde{T}{\bf g}({\vect M })$, where $\tilde P:=\mathrm{diag}(1,1,-1)$ and $\tilde T$ denote complex conjugation together with time reversal $t\to -t$~\footnote{The minus sign is a consequence of the convention
$i\dot{\vect M}={\bf f}(\vect M)$ with ${\bf f}:=i{\bf g}$ and
$\dot{\vect M}={\bf g}(\vect M)$. Since $\tilde T$ is anti-linear,
$i\to -i$, the n-PT covariance of ${\bf f}$ becomes
${\bf g}(\tilde P\tilde T\vect M)=-\tilde P\tilde T{\bf g}(\vect M)$.}.
Under the n-PT transformation, $X$ and $Y$ remain invariant, whereas $Z\to -Z$. It follows that \(Z_{*}=0\) for PT-symmetric fixed points, whereas \(Z_{*}\neq 0\) for PT-broken fixed points, where the subscript \(*\) denotes evaluation at the fixed point.

Linearization of the mean-field equations around the PT-broken fixed point $\vect{M}_{*}$ yields
\begin{align}
\dot{\delta \vect{M}} = J\,\delta \vect{M},\quad
J=\begin{pmatrix}Z_*\mathcal{E}(Z_*^2) & Z_*\mathcal{E}(Z_*^2) & Z_*^2\mathcal{E}(Z_*^2)\\
Z_*\mathcal{E}(Z_*^2) & Z_*\mathcal{E}(Z_*^2) & Z_*^2\mathcal{E}(Z_*^2)\\
\mathcal{E}(Z_*^2)   &  \mathcal{E}(Z_*^2)   & Z_*\mathcal{E}(Z_*^2)\end{pmatrix},
\label{eq:Jgeneral}  
\end{align}
where $\delta \vect{M}:=\vect{M}-\vect{M}_{*}$ is a deviation from the fixed point of continuous phase transitions, and \(\mathcal{E}(Z_*^2)\) denotes a generic even function of \(Z_*\). Different occurrences of \(\mathcal{E}(Z_*^2)\) in Eq.~\eqref{eq:Jgeneral} generally represent different functions (see Supplemental Materials Sec.~B ~\cite{supp}).
At the PT-symmetry breaking point, $Z_*=0$, the two collective linear modes generically coalesce, rendering the Jacobian non-diagonalizable and giving rise to a CEP~\cite{Nakanishi_review}; the associated coalescence direction is $\mathbf{e}_z$.

In a rotated transverse frame, one can find~\cite{supp}
\begin{align}
J^\prime_\perp=
\begin{pmatrix}
 Z_*\mathcal{E}(Z_*^2)&  \mathcal{E}(Z_*^2)\\[2pt]
 Z_*^2\mathcal{E}(Z_*^2)&  Z_*\mathcal{E}(Z_*^2)
\end{pmatrix},\ 
D^\prime_\perp=
\begin{pmatrix}
\mathcal{E}(Z_*^2)& Z_*\mathcal{E}(Z_*^2)\\[2pt]
Z_*\mathcal{E}(Z_*^2) & D_{z}+ Z_*^2\mathcal{E}(Z_*^2)
\end{pmatrix},
\label{eq:Dperp}   
\end{align}
where $J^\prime_\perp$ is the transverse Jacobian and $D^\prime_\perp$ the corresponding transverse diffusion matrix in that frame. 
The $Z$-independent part $D_z$ can be nonzero when the Lindblad operators contain an explicit $m_z$ component~\cite{supp}, for example through a dephasing channel $L_\mu \propto m_z$, whereas it vanishes if all Lindblad operators depend only on $m_x$ and $m_y$.

\noindent\textit{Universal CEP scaling of quantum fluctuations.—}
We now derive the central result of this work on a stable PT-broken fixed-point branch. Throughout this section, \(Z_*\) denotes the nonzero order parameter of the selected stable PT-broken branch, and the CEP is approached as a one-sided limit \(Z_*\to0\). The stationary Lyapunov analysis below is not applied to the PT-symmetric fixed point itself, where its fixed point is the center and the covariance generally does not converge to a time-independent stationary value~\cite{Nakanishi3,Nakanishi_review}.

Inserting the generic CEP forms of \(J'_\perp\) and \(D'_\perp\) in Eq.~\eqref{eq:Dperp} into the steady-state Lyapunov equation,
\(J'_\perp\Sigma'_\perp+\Sigma'_\perp J_\perp^{\prime T}+D'_\perp=0\),
one obtains the universal asymptotic scaling of the covariance.
When \(D_z=0\), the covariance matrix obeys
\begin{align}
\label{sigma}
\Sigma_{11}^\prime\sim |Z_*|^{-1}, \qquad
\Sigma_{22}^\prime\sim |Z_*|, \qquad
\Sigma_{12}^\prime\sim O(1),
\end{align}
near the CEP.
Equation~\eqref{sigma} shows that the quantum covariance becomes singularly anisotropic in the transverse fluctuation plane.
Consequently, the principal variances scale as
\begin{align}
\lambda_{\min}({\Sigma}^\prime_\perp)
=\xi_S^2\sim |Z_*|,\quad
\lambda_{\max}({\Sigma}^\prime_\perp)
\sim |Z_*|^{-1},
\label{eq:CEP_squeezing_scaling}
\end{align}
where $\lambda_{\rm max}$ denotes the largest eigenvalue.

The same covariance matrix also fixes the principal axes.
In the rotated frame, the coalesced vector satisfies
\(R\mathbf e_z=(-\sqrt{1-Z_*^2},0,Z_*)\), so its transverse projection is
parallel to the first transverse axis.
The eigenvector of \(\Sigma'_\perp\) associated with \(\lambda_{\max}\)
is proportional to \((1,O(Z_*))\).
Hence, as \(|Z_*|\to0\), the anti-squeezed axis locks to the coalesced mode axis.
Thus, a CEP is directly imprinted on a genuinely quantum observable: in the thermodynamic Gaussian regime, the normalized squeezed variance vanishes linearly with the order parameter, while the orthogonal variance diverges.

The condition \(D_z=0\) is essential for the asymptotic scaling in Eq.~\eqref{eq:CEP_squeezing_scaling}.
When \(D_z\neq0\), the small-\(|Z_*|\) covariance is qualitatively modified.
Instead of Eq.~\eqref{sigma}, one generically obtains
$\Sigma_{11}^\prime\sim D_z |Z_*|^{-3},\ 
\Sigma_{12}^\prime\sim D_z |Z_*|^{-2},\ 
\Sigma_{22}^\prime\sim D_z |Z_*|^{-1}$ (see Supplemental Materials Sec.~B.6 ~\cite{supp}).
The smallest principal variance is then no longer suppressed as \( |Z_*| \); generically it grows as \(D_z |Z_*|^{-1}\).
Thus any finite \(D_z\) asymptotically cuts off the parametric CEP-induced squeezing sufficiently close to the CEP.

\begin{figure*}[t]
  \centering
\includegraphics[width=0.95\linewidth]{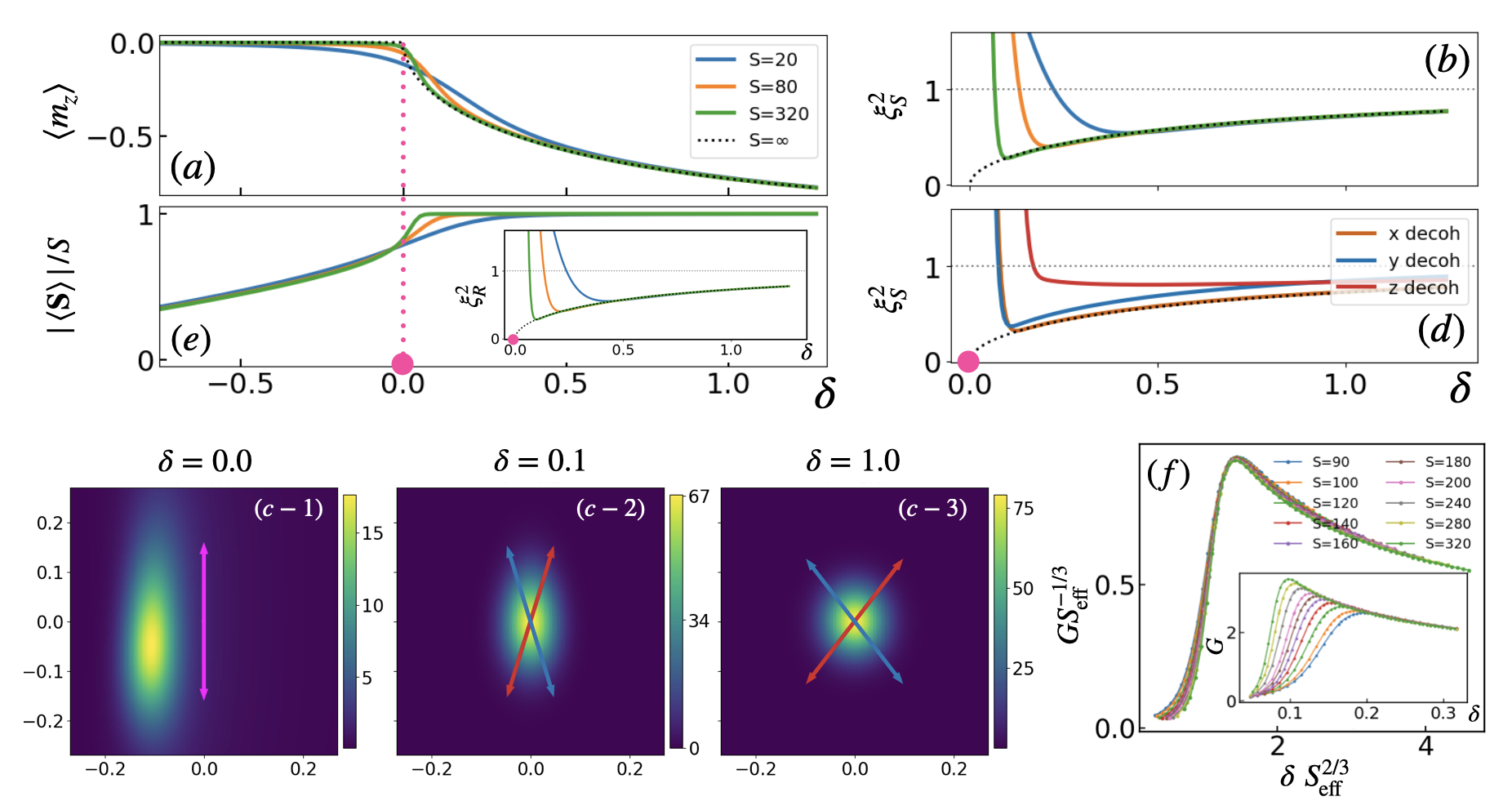}
\caption{\justifying\small Numerical calculation of the dissipative collective spin model~\eqref{geneddm}.
\textbf{(a)} Order parameter $\langle m_z\rangle$ and \textbf{(b)} Kitagawa Ueda spin-squeezing parameter $\xi_S^2$ as functions of $\delta:=(\kappa-\kappa_c)/\omega$ in the steady-state, with $g/\omega=2$ . Colored solid lines are finite-$S$ numerics, while the black dotted curve shows the large-$S$ prediction. The pink marker indicates the CEP. \textbf{(c-1)--(c-3)} Husimi $Q$ function of the steady state on the tangent plane orthogonal to the mean spin for the three values of $\delta=0,\ 0.1,\ 1.0$ for $S=500$. The two collective excitation vectors obtained from mean-field analysis become nearly parallel and coalesce at the CEP (pink).
\textbf{(d)} Spin-squeezing parameter $\xi_S^2$ in the presence of one extra dephasing channel, $L_x=\sqrt{\gamma_xS}\,m_x$ (orange), $L_y=\sqrt{\gamma_yS}\,m_y$ (blue), or $L_z=\sqrt{\gamma_z S}\,m_z$ (red) with $\gamma_\alpha/\omega=1$ and $S=320$.
The black dotted curve shows the decoherence-free reference in the thermodynamic limit. \textbf{(e)} Steady-state mean spin strength $|\braket{\bm{S}}|/S$. inset: Steady-state Wineland spin squeezing parameter.
\textbf{(f)} Log-corrected finite-size scaling collapse of the inverse squeezing
$G:=1/\xi_S^2$ near the CEP. 
The main panel plots $GS_{\rm eff}^{-1/3}$ versus
$\delta \ S_{\rm eff}^{2/3}$. 
Inset: unscaled $G$ as a function of $\delta$, showing the growth of maximum of 
$G$ and the drift of optimal $\delta$ toward the CEP with increasing $S$. 
}
\label{fig:squeezing_CEP}
\end{figure*}

\noindent\textit{Example.—}
As a concrete example, we consider the collective-spin model~\cite{Nakanishi3, Piccitto} with
\begin{align}
\label{geneddm}
H_S = S\!\left(g\,m_x + \omega\, m_z^2/2\right),\qquad
L_-=\sqrt{\kappa S}\, m_-,
\end{align}
where $m_-=m_x- i m_y$.
The parameter $g$ sets the coherent transverse field along $x$, $\omega$ controls the nonlinear one-axis-twisting interaction, and $\kappa$ is the collective decay rate associated with the jump operator $m_-$. 
This model satisfies the L-\(\mathcal{PT}\) symmetry~\eqref{HuberPT}. 
It has a CEP at \(\kappa=\kappa_c\), with \(\kappa_c:=\sqrt{g^2-\omega^2}\). 
Introducing the control parameter 
\(\delta:=(\kappa-\kappa_c)/\omega\) and \(r:=\kappa_c/\omega\), the stable PT-broken branch is \(Z_*=-\sqrt{\delta(2r+\delta)/[1+(r+\delta)^2]}\), which vanishes at the CEP, as shown in Figure~\ref{fig:squeezing_CEP}(a)~\cite{Nakanishi3}.

To assess the robustness of the CEP-enhanced steady-state squeezing against additional noise channels, we add a single collective dephasing operator $L_\alpha=\sqrt{\gamma_\alpha S}\,m_\alpha$ with $\alpha=x,y,z$.
Since these linear jump operators do not modify the mean-field equation, the CEP remains. We then obtain the explicit form of Eq.~\eqref{eq:Dperp} as
\begin{align}
J^\prime_\perp=
\begin{pmatrix}
 \kappa Z_*&  -\omega\\[2pt]
 \omega Z_*^2&  \kappa Z_*
\end{pmatrix},\quad D'_{\perp}
=
\begin{pmatrix}
d_{11}
&
d_{12}Z_*
\\[6pt]
d_{12}Z_*
&
\gamma_z+d_{22}Z_*^2
\end{pmatrix},
\end{align}
where \(d_{11}\), \(d_{12}\), and \(d_{22}\) are independent of \(Z_*\):
\(d_{11}=2\kappa+(\gamma_x\kappa^2+\gamma_y\omega^2)/(\kappa^2+\omega^2)\),
\(d_{12}=\kappa\omega(\gamma_x-\gamma_y)/(\kappa^2+\omega^2)\), and
\(d_{22}=2\kappa+(\gamma_x\omega^2+\gamma_y\kappa^2)/(\kappa^2+\omega^2)-\gamma_z\)  (see Supplemental Materials Sec.~C ~\cite{supp}). We remark that the one twisting term \(\omega m_z^2\) tends to enhance spin squeezing. The key point, however, is that the same squeezing scaling survives~\eqref{eq:CEP_squeezing_scaling} in the limit \(\omega \to 0\). Hence, this scaling is not generated solely by the one-axis twisting term~\footnote{Even at \(\omega = 0\), the same squeezing scaling in Eq.~\eqref{eq:CEP_squeezing_scaling} survives~\cite{carollo2022exact, Buonaiuto}. 
At this point, however, the projected \(2\times2\) tangent-plane dynamics no longer exhibits an exceptional point, although the full \(3\times3\) linearized matrix remains at a CEP. 
Thus, what fails at \(\omega = 0\) is only the tangent-plane geometric interpretation, not the squeezing scaling itself.}.

Figure~\ref{fig:squeezing_CEP}(b) shows the spin squeezing parameter $\xi_S^2$ defined in Eq.~\eqref{KU} in the steady-state without decoherence channels.
In the vicinity of the CEP, $\xi_S^2$ is strongly suppressed.
As $S$ increases, numerical results approach the large-$S$ analytic prediction (black dotted curve).

The CEP-induced anisotropy can be visualized in phase space through the spin Husimi-$Q$ distribution,
$Q(\theta,\phi)=\,\langle \theta,\phi|{\rho}|\theta,\phi\rangle/C$, 
with normalization constant $C:={4\pi}/({N+1})$, where $|\theta,\phi\rangle$ is the spin-coherent state pointing along the Bloch-sphere direction $(\theta,\phi)$.
Figure~\ref{fig:squeezing_CEP}(c) shows $Q$ in local Cartesian coordinates obtained by projecting the Bloch sphere onto the tangent plane at the mean-spin direction.
At the CEP, $Q$ becomes strongly anisotropic and exhibits a pronounced elongation.
Its major axis is approximately aligned with the coalescence direction of the linearized collective dynamics, indicating that the CEP strongly enhances quantum fluctuations in one transverse quadrature while suppressing them in the orthogonal one.

Figure~\ref{fig:squeezing_CEP} (d) compares the resulting steady-state squeezing $\xi_S^2$ in the presence of decoherence.
The dephasing channels $L_x$ and $L_y$ remain close to the trend of decoherence-free over a broad range, whereas the dephasing channel $L_z$ produces substantially larger $\xi_S^2$.
This behavior is consistent with our diffusion-based viewpoint, namely that noise channels generating a finite diffusion component $D_z$ along the CEP-relevant direction more efficiently suppress the CEP-induced squeezing enhancement.

Finally, we relate the CEP-induced squeezing to spin entanglement. 
While our main result concerns the Kitagawa–Ueda parameter \(\xi_S^2\), 
entanglement and metrological usefulness are certified by the Wineland parameter
\(
\xi_R^2=\frac{S^2}{|\langle \mathbf S\rangle|^2}\xi_S^2
\)~\cite{Wineland1992PRA,Kitagawa1993PRA,PezzeSmerzi2009PRL,Hyllus2012PRA,Toth2012PRA,Pezze2018RMP,Sorensen2001PRL}.
When the selected PT-broken steady state remains asymptotically polarized,
\(|\langle \mathbf S\rangle|/S\to 1\), the CEP scaling of \(\xi_S^2\) directly carries over to
\(\xi_R^2\)~\footnote{For the class~\eqref{eq:quadH_linL}, a gapped steady state admits an asymptotically exact semiclassical mean-field description in the thermodynamic limit~\cite{da2023sufficient}. 
In addition, each stable PT-broken fixed point is nondegenerate and has a finite excitation gap~\cite{Nakanishi_review}. 
Therefore, when the dynamics selects a single stable PT-broken branch, the corresponding steady state is expected to remain asymptotically polarized, \( |\braket{\bm S}|/S\to 1 \). 
In this regime, \(\xi_R^2\simeq \xi_S^2\) for large \(S\). 
Thus, CEP-induced squeezing with \(\xi_S^2<1\) also gives \(\xi_R^2<1\), witnessing spin entanglement.}. In the representative model, Figure~\ref{fig:squeezing_CEP} (e) confirms that the polarization remains close to unity and that \(\xi_R^2<1\) near the CEP, showing that the CEP-induced suppression of \(\xi_S^2\) corresponds to steady-state spin entanglement.

\noindent\textit{Finite-size scaling.—}
We consider the fully connected spin system, so finite-$S$ rounding is estimated by a Ginzburg-type argument~\cite{Rikvold1993Critical,ColonnaRomano2014Anomalous}. 
Around a PT-broken fixed point, the anti-squeezed fluctuation scales as
$\left(\lambda_{\max}(\Sigma_\perp ^\prime)/S\right)^{1/2}
\sim
(S|Z_*|)^{-1/2}$.
Requiring this fluctuation to remain smaller than the order-parameter displacement $|Z_*|$ gives
$S|Z_*|^3=O(1)$.
Together with the mean-field relation \(|Z_*|\sim\delta^{1/2}\)~\cite{Nakanishi_review,supp}, this yields 
$Z_S(\delta)
=
S^{-1/3}{\cal Z}(S^{2/3}\delta),
\ 
{\cal Z}(x)\sim x^{1/2}\ (x\gg1)$,
as numerically verified in Supplementary Materials 
Sec.C~\cite{supp}.

Because \(\xi_S^2=\lambda_{\min}(\Sigma'_\perp)\sim |Z_*|\), the same criterion gives the algebraic cutoff
\((\xi_S^2)_{\rm cutoff}\sim S^{-1/3}\). 
This exponent is consistent with previous results for the \(\omega=0\) case~\cite{SomechShahmoon2024}, although the Ginzburg argument does not fix possible marginal logarithmic corrections. 
Motivated by the logarithmic corrections found for \(\omega=0\)~\cite{BarberenaRey2024}, we introduce \(S_{\rm eff}:=S/\log S\) and test the scaling form
\begin{align}
G(\delta; S)
=
S_{\rm eff}^{1/3}
{\cal G}\!\left(\delta S_{\rm eff}^{2/3}\right),
\quad
G:=1/\xi_S^2 .
\label{eq:log_corrected_fss}
\end{align}

Figure~\ref{fig:squeezing_CEP} (f) shows this collapse by plotting
$G(\delta; S) S_{\rm eff}^{-1/3}$ against $\delta\ S_{\rm eff}^{2/3}$.
The inset displays the unscaled data, where the enhancement of $G$ grows and shifts toward the CEP as $S$ increases. 
In the rescaled plot, the curves for different $S$ fall onto a common scaling function, including the location and height of the maximum. 
This agreement supports the logarithmically corrected finite-size scaling in Eq.~\eqref{eq:log_corrected_fss}.

\noindent\textit{Implementation.—}
The CEP-induced squeezing mechanism is not an artifact of purely collective-spin models. 
It also appears in a spin-boson realization in which a collective spin is coupled to a lossy cavity mode,
\begin{align}
H = H_{\rm S} + \lambda \left(a S_+ + a^\dagger S_-\right)/\sqrt{N}, 
\quad
L = \sqrt{\kappa}\, a,
\end{align}
where \(N\) is the number of spins, \(a\) is the cavity annihilation operator (see Supplemental Materials Sec.~D ~\cite{supp}).
In this model, the mean-field dynamics exhibits a CEP, and the corresponding steady-state fluctuations show the same CEP-controlled anisotropic structure.
Such a spin-boson structure is natural in cavity-QED systems, although realizing the CEP considered here requires an appropriate spin Hamiltonian \(H_{\rm S}\) and parameter regime.
A closely related platform has been realized with cold \(^{88}\)Sr atoms in a high-finesse optical cavity, where \(N\sim10^3\text{--}10^4\) atoms are collectively coupled to the cavity and both the cavity output field and collective-spin observables are experimentally accessible~\cite{Song2025SciAdv}.
While that experiment is not a direct implementation of the present CEP-squeezing model, it already provides the essential ingredients needed to explore CEP-induced steady-state spin squeezing after suitable Hamiltonian engineering.

\noindent\textit{Conclusion.—}
We have shown that CEPs provide a mechanism for generating steady-state quantum spin squeezing in open collective-spin systems. 
Previous open-system approaches have generated spin squeezing by selecting or stabilizing squeezed collective states through bosonic-reservoir engineering~\cite{groszkowski2022prx}, cavity-mediated collective dark-state preparation~\cite{dallaTorre2013prl}, or postselected no-jump non-Hermitian dynamics~\cite{LeeReiterMoiseyev2014PRL}. 

By contrast, the present squeezing is not imposed by an engineered target state, a dark-state condition, or postselected spectral filtering, but emerges from the critical Jordan-block structure of a CEP in an unconditional GKSL steady state. 
More broadly, this covariance-geometric viewpoint may also connect to squeezing near zero-frequency soft modes in isolated Hamiltonian systems, where analogous nilpotent structures can appear in effective first-order dynamical representations~\footnote{A similar nilpotent matrix can appear in the semiclassical phase-space description of an isolated Hamiltonian collective-spin system, or in the bosonic Bogoliubov dynamical matrix of a Hermitian quadratic Hamiltonian~\cite{Flynn2020QuadraticBosons,Xie2021HopfieldBogoliubov,Yamamoto2025MagnonBEC}. This reflects zero-frequency softening of a Hermitian system in an effective first-order dynamical representation, not an exceptional point of the microscopic generator of unitary dynamics, and is therefore distinct from the open-system CEPs studied here. }. Clarifying whether this viewpoint can be extended to such isolated-system soft-mode squeezing is an interesting direction for future work.

Our results identify CEPs, within this class of open collective-spin systems, as universal organizing structures for the covariance geometry of steady-state quantum fluctuations, with spin squeezing serving as a direct quantum signature of CEP-controlled correlations.

\begin{acknowledgments}
\textit{Acknowledgments.—} We thank Ryo Hanai, Shohei Imai, and Taiga Nakamoto for helpful discussions. This work was supported by JSPS KAKENHI, Grant No.24K22850.
\end{acknowledgments}

\makeatletter
\immediate\write\@auxout{\string\citation{apsrev42Control}}
\makeatother
\nocite{carollo2022exact,Buonaiuto,da2023sufficient,Flynn2020QuadraticBosons,Xie2021HopfieldBogoliubov,Yamamoto2025MagnonBEC}
\bibliographystyle{apsrev4-2}
\bibliography{0ref_cep_squeezing_arxiv_fixed}

\onecolumngrid
\newpage
\begin{center}
{\large \textbf{Supplemental Materials: Quantum Spin Squeezing Enhanced by Critical Exceptional Points}}\\[4pt]
\end{center}

\makeatletter
\def\@appendixcntformat#1{\csname the#1\endcsname.}%
\def\@hangfrom@appendix#1#2#3{%
  #1\@if@empty{#2}{#3}{#2\@if@empty{#3}{}{\quad #3}}%
}%
\makeatother

\makeatletter
\newcounter{appsec}
\newcounter{appsub}[appsec]
\renewcommand{\theappsec}{\Alph{appsec}}
\renewcommand{\theappsub}{\theappsec.\arabic{appsub}}

\newcommand{\AppSection}[1]{%
  \refstepcounter{appsec}%
  \setcounter{equation}{0}\renewcommand{\theequation}{\theappsec.\arabic{equation}}%
  \setcounter{figure}{0}\renewcommand{\thefigure}{\theappsec.\arabic{figure}}%
  \setcounter{table}{0}\renewcommand{\thetable}{\theappsec.\arabic{table}}%
  \section*{\theappsec.\ #1}%
}

\newcommand{\AppSubsection}[1]{%
  \refstepcounter{appsub}%
  \subsection*{\theappsub.\ #1}%
}
\makeatother

\AppSection{Derivation of the Lyapunov equation for spin-boson systems}

\AppSubsection{Setup and large-\(N\) scaling}

We consider \(N\) identical spin-\(1/2\) particles coupled to a single bosonic cavity mode \(a\).
The collective-spin operators are
\begin{equation}
S_\alpha = \frac12 \sum_{i=1}^N \sigma_i^\alpha,
\qquad
\alpha \in \{x,y,z\},
\end{equation}
and we introduce the intensive variables
\begin{equation}
m_\alpha := \frac{2}{N} S_\alpha,
\qquad
b := \frac{a}{\sqrt N}.
\end{equation}
With this normalization, the commutators are explicitly of order \(N^{-1}\):
\begin{equation}
[m_\alpha,m_\beta]
=
\frac{2i}{N}\sum_\gamma \varepsilon_{\alpha\beta\gamma} m_\gamma,
\qquad
[b,b^\dagger] = \frac1N,
\qquad
[m_\alpha,b]=[m_\alpha,b^\dagger]=0.
\label{eq:scaled_comm}
\end{equation}
We also use the bosonic quadratures
\begin{equation}
q = \frac{1}{\sqrt2}(b+b^\dagger),
\qquad
p = \frac{i}{\sqrt2}(b-b^\dagger),
\end{equation}
and collect the Hermitian intensive operators into
\begin{equation}
\bm{x} := (m_x,m_y,m_z,q,p)^T.
\end{equation}

The dynamics is governed by the GKSL master equation
\begin{equation}
\dot\rho
=
\hat{\mathcal L}[H,\{L_\mu\}] \rho
=
-i[H,\rho]
+
\sum_\mu
\left(
L_\mu \rho L_\mu^\dagger
-
\frac12\{L_\mu^\dagger L_\mu,\rho\}
\right).
\label{eq:GKSL_general}
\end{equation}
We assume the standard fully connected scaling
\begin{equation}
H = N h(\bm{x}),
\qquad
L_\mu = \sqrt N\, \ell_\mu(\bm{x}),
\label{eq:H_extensive}
\end{equation}
with \(h(\bm{x})\) at most quadratic and \(\ell_\mu(\bm{x})\) linear in the intensive variables. Under this assumption, fluctuations of \(x_i\) are of order \(N^{-1/2}\), and the Gaussian description closes at leading order in \(1/N\)~\cite{CarolloLesanovsky2024PRL,DuboisSaalmannRost2021,Fiorelli2024JPhysA}.

For any observable \(O\), the evolution of its expectation value is generated by the adjoint Lindbladian,
\begin{equation}
\frac{d}{dt}\langle O\rangle = \langle \hat{\mathcal L}^\dagger[O]\rangle,
\qquad
\hat{\mathcal L}^\dagger[O]
=
i[H,O]
+
\sum_\mu
\left(
L_\mu^\dagger O L_\mu
-
\frac12\{L_\mu^\dagger L_\mu,O\}
\right).
\label{eq:adjoint}
\end{equation}

\AppSubsection{Action of GKSL generators on operator products}

The Hamiltonian part obeys the usual product rule,
\[
\hat{\mathcal H}^\dagger[AB]
=
\hat{\mathcal H}^\dagger[A]\,B
+
A\,\hat{\mathcal H}^\dagger[B].
\]
For the dissipative part, an additional commutator term appears. A direct expansion gives
\begin{equation}
\hat{\mathcal L}^\dagger[AB]
=
\hat{\mathcal L}^\dagger[A]\,B
+
A\,\hat{\mathcal L}^\dagger[B]
+
\sum_\mu [L_\mu^\dagger,A][B,L_\mu].
\label{eq:GKSL_product_rule}
\end{equation}
This identity is the key ingredient in the derivation of the covariance equation.

\AppSubsection{First moments and covariance matrix}

We define the first moments
\begin{equation}
X_i := \langle x_i\rangle,
\qquad
\bm{X} := \langle \bm{x}\rangle,
\end{equation}
and the fluctuations
\begin{equation}
\Delta x_i := x_i - X_i,
\qquad
\langle \Delta x_i\rangle = 0.
\end{equation}
To obtain an \(O(1)\) covariance matrix in the large-\(N\) limit, we introduce the scaled fluctuations
\begin{equation}
\eta_i := \sqrt N\, \Delta x_i
\end{equation}
and the symmetrized covariance matrix
\begin{equation}
\Sigma_{ij}
:=
\frac12 \langle \eta_i \eta_j + \eta_j \eta_i\rangle
=
\frac N2 \langle \{\Delta x_i,\Delta x_j\}\rangle.
\label{eq:cov_scaled_def}
\end{equation}
Equivalently, the symmetrized second moments satisfy
\begin{equation}
X_{ij}
:=
\frac12 \langle \{x_i,x_j\}\rangle
=
X_i X_j + \frac{1}{N} \Sigma_{ij}.
\label{eq:M_split}
\end{equation}

The exact evolution of the first moments is
\begin{equation}
\dot X_i = \langle \hat{\mathcal L}^\dagger[x_i]\rangle.
\label{eq:first_moment_exact}
\end{equation}
Applying Eq.~\eqref{eq:GKSL_product_rule} to \(A=x_i\) and \(B=x_j\), and then symmetrizing, yields the exact evolution of the second moments,
\begin{align}
\dot X_{ij}
=
\frac12 \left\langle \{\hat{\mathcal L}^\dagger[x_i],x_j\}\right\rangle
+
\frac12 \left\langle \{x_i,\hat{\mathcal L}^\dagger[x_j]\}\right\rangle
+
\frac12 \sum_\mu
\left\langle
[L_\mu^\dagger,x_i][x_j,L_\mu]
+
[L_\mu^\dagger,x_j][x_i,L_\mu]
\right\rangle.
\label{eq:dotM_exact_general}
\end{align}

\AppSubsection{Lyapunov equation in the large-\(N\) limit}

To extract the leading large-\(N\) dynamics, we now expand around the mean field. We define the drift vector and Jacobian by
\begin{equation}
f_i(\bm{X})
:=
\left.\langle \hat{\mathcal L}^\dagger[x_i]\rangle\right|_{\mathrm{mf}},
\qquad
J_{ij}(\bm{X})
:=
\frac{\partial f_i(\bm{X})}{\partial X_j},
\label{eq:drift_and_jacobian}
\end{equation}
where \(\big|_{\mathrm{mf}}\) denotes factorization of operator products, \(\langle x_i x_j\rangle \to X_i X_j\).

For the class~\eqref{eq:H_extensive}, since \(\Delta x_i = O(N^{-1/2})\), the adjoint dynamics of \(x_i\) can be expanded as
\begin{equation}
\hat{\mathcal L}^\dagger[x_i]
=
f_i(\bm{X})
+
\sum_k J_{ik}(\bm{X})\, \Delta x_k
+
O(\Delta x^2).
\label{eq:drift-expansion}
\end{equation}
The \(O(\Delta x^2)\) term is \(O(N^{-1})\), so at leading nontrivial order only the linear fluctuation term contributes to the covariance dynamics.

Substituting Eq.~\eqref{eq:drift-expansion} into the exact second-moment equation and retaining the leading terms in \(1/N\), one obtains a closed Lyapunov-type equation for the covariance matrix:
\begin{equation}
\dot \Sigma
=
J(\bm{X})\,\Sigma
+
\Sigma\,J(\bm{X})^T
+
D(\bm{X})
+
O(N^{-1}),
\label{eq:lyapunov_form}
\end{equation}
where the diffusion matrix is
\begin{equation}
D_{ij}(\bm{X})
:=
\frac{N^2}{2}
\sum_\mu
\left.
\left\langle
[\ell_\mu^\dagger,x_i][x_j,\ell_\mu]
+
[\ell_\mu^\dagger,x_j][x_i,\ell_\mu]
\right\rangle
\right|_{\mathrm{mf}}.
\label{eq:diffusion_matrix_def}
\end{equation}

Equation~\eqref{eq:lyapunov_form} is the central result of this section: under the quadratic-linear scaling in Eq.~\eqref{eq:H_extensive}, the first moments and the covariance matrix form a closed set at leading order in \(1/N\).

\newpage

\AppSection{The diffusion matrix and Jacobian for L-$\mathcal{PT}$ symmetric collective spin models}
In this section, we consider the collective spin model eliminating bosonic degrees of freedom (e.g. in the bad cavity limit).
When the model has an L-$\mathcal{PT}$ symmetry,
if Lindblad operators $L_\mu$ is present,
its partner
\begin{equation}
\mathbb{PT}(L_\mu) := PT\,L_\mu^\dagger\,(PT)^{-1}
\end{equation}
is also present in the set of Lindblad operators.
Throughout this section, we view mean-field expectations as functions of the steady-state
spin direction $M=(X,Y,Z):=(X_1,X_2,X_3)$ evaluated on a fixed-point branch.

\AppSubsection{Diffusion matrix}
For any normalized Lindblad operator $l$, we define
\begin{equation}
\Gamma_{ij}[l]
:=2S^2 \left([l^\dagger,x_i]\,[x_j,l] + [l^\dagger,x_j]\,[x_i,l]\right),
\label{eq:Gamma_def}
\end{equation}
which is the order $\mathcal{O}(1)$. Here, $N=2S$.
Then, the diffusion matrix with mean-field approximation~\eqref{eq:diffusion_matrix_def} can be written as
\begin{equation}
D_{ij}(X,Y,Z)
= \sum_\mu \Big\langle \Gamma_{ij}[l_\mu]\Big\rangle\Big|_{\rm mf}(X,Y,Z),
\label{eq:D_from_Gamma}
\end{equation}
where $\Big|_{\rm mf}$ denotes the mean-field approximation.
Grouping the L-$\mathcal{PT}$ pairs $(L_\mu,\mathbb{PT}(L_\mu))$ gives
\begin{equation}
D_{ij}(X,Y,Z)
= \sum_\mu
\left[
\Big\langle \Gamma_{ij}[l_\mu]\Big\rangle\Big|_{\rm mf}(X,Y,Z)
+\Big\langle \Gamma_{ij}[\mathbb{PT}(l_\mu)]\Big\rangle\Big|_{\rm mf}(X,Y,Z)
\right].
\label{eq:D_pair_sum}
\end{equation}

At leading order in $1/N$, the following equation satisfies
\begin{equation}
\Gamma_{ij}[\mathbb{PT}(l_\mu)]
=
p_{ij}\,\hat{\mathcal{P}}\hat{\mathcal{T}}\!\left(\Gamma_{ij}[l_\mu]\right)
+\mathcal{O}(1/S),
\label{eq:Gamma_PT_relation}
\end{equation}
where $p_{ij}=+1$ for $ij=xx,yy,xy,zz$ and $p_{ij}=-1$ for $ij=xz,yz$ (and by symmetry also for $zx,zy$).
On mean-field arguments, $\mathcal{PT}$ superoperator acts as
\begin{equation}
\hat{\mathcal{P}}\hat{\mathcal{T}}\big(F(m_x,m_y,m_z)\big)=F(m_x,m_y,-m_z),
\end{equation}
so that
\begin{equation}
\Big\langle \Gamma_{ij}[\mathbb{PT}(l_\mu)]\Big\rangle\Big|_{\rm mf}(X,Y,Z)
=
p_{ij}\,\Big\langle \Gamma_{ij}[l_\mu]\Big\rangle\Big|_{\rm mf}(X,Y,-Z).
\label{eq:Gamma_expect_Zflip}
\end{equation}
Substituting \eqref{eq:Gamma_expect_Zflip} into \eqref{eq:D_pair_sum} yields
\begin{equation}
D_{ij}(X,Y,Z)=\sum_\mu
\left[
\Big\langle \Gamma_{ij}[l_\mu]\Big\rangle\Big|_{\rm mf}(X,Y,Z)
+p_{ij}\,\Big\langle \Gamma_{ij}[l_\mu]\Big\rangle\Big|_{\rm mf}(X,Y,-Z)
\right].
\end{equation}
Hence, the lab-frame components obey the $Z$-parity constraint
\begin{equation}
D_{xx},\,D_{yy},\,D_{xy},\,D_{zz}\ \text{are even in $Z$},\qquad
D_{xz},\,D_{yz}\ \text{are odd in $Z$}.
\label{eq:D_parity_components}
\end{equation}
Therefore, it can be written as
\begin{equation}
D(X,Y,Z)=
\begin{pmatrix}
\mathcal{E}(Z^2) & \mathcal{E}(Z^2) & Z\times\mathcal{E}(Z^2)\\
\mathcal{E}(Z^2) & \mathcal{E}(Z^2) & Z\times \mathcal{E}(Z^2)\\
Z\times\mathcal{E}(Z^2)& Z\times\mathcal{E}(Z^2) & \mathcal{E}(Z^2)
\end{pmatrix},
\label{eq:D_general_withSz}
\end{equation}
where \(\mathcal{E}(Z^2)\) denotes a generic even function of \(Z\), that is, a function of \(Z^2\). Different occurrences of \(\mathcal{E}(Z^2)\) generally represent different functions. 

\subsubsection{Stronger power counting when no jump contains $m_z$.}
We assume that every Lindblad operator has no explicit $m_z$ component,
\begin{equation}
l_\mu = l_\mu(m_x,m_y).
\label{eq:no_Sz_in_jumps}
\end{equation}
Then $[m_x,l_\mu]$ and $[m_y,l_\mu]$ necessarily carry one factor of $m_z$ (through $[m_x,m_y]=im_z/S$), so  
\begin{equation}
D_{xx},\,D_{yy},\,D_{xy}\ \ \text{are even in $Z$}.
\end{equation}

As a result, the leading-order
lab-frame diffusion tensor admits the schematic structure
\begin{equation}
D(X,Y,Z)=
\begin{pmatrix}
Z^2\times \mathcal{E}(Z^2) & Z^2\times \mathcal{E}(Z^2) & Z \times\mathcal{E}(Z^2)\\
Z^2\times \mathcal{E}(Z^2) & Z^2 \times\mathcal{E}(Z^2)& Z \times\mathcal{E}(Z^2)\\
Z\times \mathcal{E}(Z^2)& Z\times\mathcal{E}(Z^2)& \mathcal{E}(Z^2)
\end{pmatrix}.
\label{eq:D_structure_noSz}
\end{equation}
As an example, for collective decay with $l=m_-:=m_x-im_y$ one finds
\begin{equation}
D(X,Y,Z)=
\begin{pmatrix}
Z^2 & 0 & -XZ\\
0 & Z^2 & -YZ\\
-XZ & -YZ & X^2+Y^2
\end{pmatrix}.
\end{equation}

\AppSubsection{\(Z\)-parity of the Jacobian}
\label{sec:Zparity_J_general}

At the mean-field level, L-\(\mathcal{PT}\) symmetry implies the nonlinear
\(\mathcal{PT}\) covariance
\begin{equation}
\bm g(\tilde P\bm M)=-\tilde P\,\bm g(\bm M).
\label{eq:nPT_covariance_general}
\end{equation}
For the \(\mathcal{PT}\) transformation
\(\mathcal{PT}=\prod_i\sigma_x^i K\), one has
\[
\tilde P=\mathrm{diag}(p_x,p_y,p_z),\qquad
p_x=p_y=+1,\quad p_z=-1 .
\]
Thus Eq.~\eqref{eq:nPT_covariance_general} gives the component-wise
parity constraints
\begin{equation}
g_x(X,Y,-Z)=-g_x(X,Y,Z),\qquad
g_y(X,Y,-Z)=-g_y(X,Y,Z),\qquad
g_z(X,Y,-Z)=+g_z(X,Y,Z).
\label{eq:g_parity_general}
\end{equation}
Hence \(g_x\) and \(g_y\) are odd functions of \(Z\), whereas \(g_z\)
is even.

We next derive the corresponding constraint on the Jacobian
\(J_{\alpha\beta}:=\partial g_\alpha/\partial M_\beta\).
Differentiating Eq.~\eqref{eq:nPT_covariance_general} with respect to
\(M_\beta\) gives
\begin{equation}
p_\beta J_{\alpha\beta}(\tilde P\bm M)
=
-p_\alpha J_{\alpha\beta}(\bm M),
\end{equation}
or equivalently
\begin{equation}
J_{\alpha\beta}(X,Y,-Z)
=
-\,p_\alpha p_\beta\,J_{\alpha\beta}(X,Y,Z).
\label{eq:J_component_parity_general}
\end{equation}
Therefore,
\begin{equation}
\begin{aligned}
&J_{xx},J_{xy},J_{yx},J_{yy},J_{zz}
\quad \text{are odd in \(Z\)},\\
&J_{xz},J_{yz},J_{zx},J_{zy}
\quad \text{are even in \(Z\)} .
\end{aligned}
\label{eq:J_parity_list_general}
\end{equation}

Assuming local analyticity in \(Z\), the most general lab-frame form
compatible with Eq.~\eqref{eq:J_parity_list_general} is
\begin{equation}
J(X,Y,Z)=
\begin{pmatrix}
Z\,\times \mathcal{E}(Z^2) & Z\times \mathcal{E}(Z^2) &  J_{xz}(Z^2) \\
Z\times \mathcal{E}(Z^2) & Z\times \mathcal{E}(Z^2) &  J_{yz}(Z^2) \\
\mathcal{E}(Z^2) & \mathcal{E}(Z^2)  & Z\times \mathcal{E}(Z^2)
\end{pmatrix},
\label{eq:J_analytic_general_form}
\end{equation}
where \(\mathcal E\)
and \(J_{\alpha z}\) are
generic analytic even functions of \(Z\). Different occurrences denote
different functions.

The fixed-point condition further constrains the apparently \(O(1)\)
entries \(J_{xz}\) and \(J_{yz}\) on a PT-broken branch. Since
\(g_x\) and \(g_y\) are odd in \(Z\), we can write
\begin{equation}
g_a(X,Y,Z)=Z\,h_a(X,Y,Z^2),\qquad a=x,y,
\label{eq:ga_odd_decomposition}
\end{equation}
with \(h_a\) analytic in its arguments. Taking the derivative with
respect to \(Z\), we obtain
\begin{equation}
J_{az}
=
\frac{\partial g_a}{\partial Z}
=
h_a(X,Y,Z^2)
+
2Z^2\,\partial_{Z^2}h_a(X,Y,Z^2),
\qquad a=x,y .
\label{eq:Jaz_general}
\end{equation}
Now, we evaluate this expression at a PT-broken fixed point
\(\bm M_*=(X_*,Y_*,Z_*)\) with \(Z_*\neq0\). The fixed-point condition
\(g_a(\bm M_*)=0\) implies
\begin{equation}
h_a(X_*,Y_*,Z_*^2)=0,\qquad a=x,y .
\end{equation}
Therefore Eq.~\eqref{eq:Jaz_general} reduces to
\begin{equation}
J_{az}(\bm M_*)
=
2Z_*^2
\left.
\partial_{u}h_a(X_*,Y_*,u)
\right|_{u=Z_*^2}
=
Z_*^2\,\mathcal E(Z_*^2),
\qquad a=x,y .
\label{eq:Jaz_fixed_branch}
\end{equation}
Thus, on the PT-broken fixed-point branch,
\begin{equation}
J_{xz}(\bm M_*)=Z_*^2\,\mathcal E(Z_*^2),
\qquad
J_{yz}(\bm M_*)=Z_*^2\,\mathcal E(Z_*^2).
\label{bxz}
\end{equation}
Here the smooth branch dependence of \(X_*\) and \(Y_*\) has been
absorbed into the generic even functions \(\mathcal E(Z_*^2)\).

Therefore,
the Jacobian evaluated on a PT-broken fixed-point branch takes the
schematic form
\begin{equation}
J(\bm M_*)=
\begin{pmatrix}
Z_*\mathcal E(Z_*^2) & Z_*\mathcal E(Z_*^2) & Z_*^2\mathcal E(Z_*^2)\\
Z_*\mathcal E(Z_*^2) & Z_*\mathcal E(Z_*^2) & Z_*^2\mathcal E(Z_*^2)\\
\mathcal E(Z_*^2)   & \mathcal E(Z_*^2)   & Z_*\mathcal E(Z_*^2)
\end{pmatrix}.
\label{eq:J_fixed_branch_general_form}
\end{equation}

\AppSubsection{Jacobian and Jordan form}
At the transition point \(Z_*=0\), Eq.~\eqref{eq:J_fixed_branch_general_form} reduces to
\begin{equation}
J_0=
\begin{pmatrix}
0&0&0\\
0&0&0\\
\mu&\nu&0
\end{pmatrix}.
\label{eq:J0_general}
\end{equation}
We assume the generic case \((\mu,\nu)\neq(0,0)\). 
Then \(J_0\) is a nonzero nilpotent matrix satisfying \(J_0^2=0\) and \(\operatorname{rank}J_0=1\). 
Therefore its Jordan form contains one nontrivial Jordan block of size two at eigenvalue zero, together with one additional zero eigenvalue. 
Moreover,
\begin{equation}
\operatorname{Im}J_0=\mathrm{span}\{\mathbf e_z\},
\end{equation}
so \(\mathbf e_z\) is the eigenvector belonging to the nontrivial Jordan block. 
Equivalently, one may choose a generalized eigenvector \(\mathbf v_1\) satisfying \(J_0\mathbf v_1=\mathbf e_z\). 
Thus the defective eigenvector, namely the coalescing direction at the CEP, is
\begin{equation}
\mathbf v_0=\mathbf e_z.
\end{equation}

\AppSubsection{Steady-state aligned frame}
Let the steady-state direction be the unit vector $\boldsymbol{M}_{\mathrm{*}}=(X,Y,Z)$ with
$X^2+Y^2+Z^2=1$. We define
\begin{equation}
r_\perp:=\sqrt{X^2+Y^2},\qquad c:=\frac{X}{r_\perp},\qquad s:=\frac{Y}{r_\perp}.
\end{equation}
A convenient orthogonal rotation that maps $\boldsymbol{M}_{\mathrm{*}}$ to the new $z'$ axis is
\begin{equation}
R=
\begin{pmatrix}
Zc & Zs & -r_\perp\\
-s & c & 0\\
r_\perp c & r_\perp s & Z
\end{pmatrix},
\qquad
R\,\boldsymbol{M}_{\mathrm{*}}=(0,0,1).
\label{eq:R_align_ss}
\end{equation}
The diffusion tensor in the aligned frame is
\begin{equation}
D'(X,Y,Z)=R\,D(X,Y,Z)\,R^{\mathsf T}.
\label{eq:Dprime_def}
\end{equation}

Provided that all Lindblad operators \(l_\mu\) obey Eq.~\eqref{eq:no_Sz_in_jumps}, the $Z$-scaling of the rotated diffusion matrix $D'$ follows directly from Eq.~\eqref{eq:D_structure_noSz} and the transformation \eqref{eq:Dprime_def}. Schematically, one finds
\begin{equation}
D'(Z)\sim
\begin{pmatrix}
\mathcal{E}(Z^2) & Z\times\mathcal{E}(Z^2) & Z\times\mathcal{E}(Z^2)\\
Z\times\mathcal{E}(Z^2) & Z^2\times\mathcal{E}(Z^2) & Z^2\times\mathcal{E}(Z^2)\\
Z\times\mathcal{E}(Z^2) & Z^2\times\mathcal{E}(Z^2) & Z^2\times\mathcal{E}(Z^2)
\end{pmatrix},
\label{eq:Dprime_structure_noSz}
\end{equation}

Similarly, transforming the Jacobian as
\begin{equation}
J'(X,Y,Z)=R\,J(X,Y,Z)\,R^{\mathsf T},
\label{eq:Jprime_def_general}
\end{equation}
it can be expressed as
\begin{equation}
J'(Z_{*})=R\,J(Z_{*})\,R^{\mathsf T}
=
\begin{pmatrix}
Z_{*}\times\mathcal{E}(Z_*^2) & \mathcal{E}(Z_*^2)  & \mathcal{E}(Z_*^2) \\
Z_{*}^2\times\mathcal{E}(Z_*^2) & Z_{*}\times\mathcal{E}(Z_*^2) & Z_{*}\times\mathcal{E}(Z_*^2)\\
Z_{*}^2\times\mathcal{E}(Z_*^2) & Z_{*}\times\mathcal{E}(Z_*^2) & Z_{*}\times\mathcal{E}(Z_*^2)
\end{pmatrix}
\label{eq:Jprime_steady_structure_noSz}
\end{equation}
for the PT-broken fixed point.

\AppSubsection{Steady-state transverse covariance from the leading $2\times2$ block}
As stated below, for spin squeezing we only need the transverse $2\times2$ block (the first two components in this frame) as
\begin{equation}
J_{\perp}(Z_{*})\equiv
\begin{pmatrix}
Z_{*}\times\mathcal{E}(Z_*^2) & \mathcal{E}(Z_*^2) \\
Z_{*}^2\times\mathcal{E}(Z_*^2) & Z_{*}\times\mathcal{E}(Z_*^2)\\
\end{pmatrix},\qquad
D_\perp(Z_{*})=
\begin{pmatrix}
\mathcal{E}(Z_*^2) & Z_{*}\times\mathcal{E}(Z_*^2)\\
Z_{*}\times\mathcal{E}(Z_*^2) & Z_{*}^2\times\mathcal{E}(Z_*^2)
\end{pmatrix}.
\label{eq:def_A_Sigma_perp}
\end{equation}
Then, the steady-state transverse covariance is determined by the reduced $2\times2$ Lyapunov equation
\begin{equation}
J_{\perp}(Z_{*})\,\Sigma_\perp(Z_{*})+\Sigma_\perp(Z_{*})\,J_{\perp}(Z_{*})^{\top}+D_\perp(Z_{*})=0.
\label{eq:Lyapunov_2x2}
\end{equation}
The $Z_{*}$ dependence of $\Sigma_\perp$ can be written schematically as
\begin{align}
\Sigma_\perp(Z_{*})\sim
\begin{pmatrix}
Z_{*}^{-1}\times\mathcal{E}(Z_*^2) & \mathcal{E}(Z_*^2)\\
\mathcal{E}(Z_*^2) & Z_{*}\times\mathcal{E}(Z_*^2)
\end{pmatrix}.
\label{sigmaZ}
\end{align}

\subsubsection{Principal variances and directions of $\Sigma_\perp$.}

Since the main text already identified the squeezing parameter with the smaller eigenvalue of $\Sigma_\perp$, the scaling form of $\Sigma_\perp$ immediately gives
\begin{equation}
\lambda_{\max}(\Sigma_\perp)\sim |Z_*|^{-1},
\qquad
\lambda_{\min}(\Sigma_\perp)\sim |Z_*|,
\end{equation}
so that the anti-squeezed variance diverges as $|Z_*|^{-1}$, whereas the squeezed one vanishes linearly in $|Z_*|$.

The corresponding principal directions are also fixed near the CEP. 
In the transverse basis in which the first coordinate is the coalescing direction, the eigenvector equation for the same scaling form of $\Sigma_\perp$ gives
\begin{equation}
\bm n_{\mathrm{asq}}
\propto
\begin{pmatrix}
1\\
Z_*\times\mathcal{E}(Z_*^2)
\end{pmatrix},
\qquad
\bm n_{\mathrm{sq}}
\propto
\begin{pmatrix}
- Z_*\times\mathcal{E}(Z_*^2)\\
1
\end{pmatrix}.
\label{eq:principal_directions_short}
\end{equation}
Hence, as $Z_*\to0$,
\begin{equation}
\bm n_{\mathrm{asq}}\to
\begin{pmatrix}
1\\
0
\end{pmatrix},
\qquad
\bm n_{\mathrm{sq}}\to
\begin{pmatrix}
0\\
1
\end{pmatrix}.
\end{equation}
Therefore, the anti-squeezed direction locks to the coalescing direction, while the squeezed direction approaches the orthogonal transverse direction.

\AppSubsection{Effect of a finite $D_z$}
If the condition in Eq.~\eqref{eq:no_Sz_in_jumps} is violated, the
transformed diffusion matrix generically takes the form
\begin{equation}
D'(Z)\sim
\begin{pmatrix}
\mathcal{E}(Z^2) & Z\times\mathcal{E}(Z^2) & Z\times\mathcal{E}(Z^2)\\
Z\times\mathcal{E}(Z^2) & D_z+Z^2\times\mathcal{E}(Z^2) & \mathcal{E}(Z^2)\\
Z\times\mathcal{E}(Z^2) & \mathcal{E}(Z^2) & \mathcal{E}(Z^2)
\end{pmatrix},
\label{eq:Dprime_structure_withSz}
\end{equation}
where $D_z$ is independent of $Z$.
The transverse block then takes the form
\begin{equation}
D_\perp(Z_*) \sim
\begin{pmatrix}
\mathcal{E}(Z_*^2) & Z_*\times\mathcal{E}(Z_*^2)\\
Z_*\times\mathcal{E}(Z_*^2) & D_z + Z_*^2\times\mathcal{E}(Z_*^2)
\end{pmatrix}.
\label{eq:Dperp_structure_constDz}
\end{equation}
This qualitatively changes the solution of the reduced Lyapunov equation. Since
\[
(J_\perp)_{21}\sim Z_*^2\times\mathcal{E}(Z_*^2),
\qquad
(J_\perp)_{22}\sim Z_*\times\mathcal{E}(Z_*^2),
\]
the \((2,2)\) component can balance the finite term \(D_z\) only if
\begin{equation}
(\Sigma_\perp^\prime)_{22} \sim D_z\, |Z_*|^{-1}.
\end{equation}
Substituting this scaling into the remaining components then gives
\begin{equation}
(\Sigma_\perp^\prime)_{12}\sim D_z\, |Z_*|^{-2},
\qquad
(\Sigma_\perp^\prime)_{11}\sim D_z\, |Z_*|^{-3}.
\end{equation}
Hence
\begin{equation}
\Sigma_\perp^\prime(Z_*) \sim D_z
\begin{pmatrix}
Z_*^{-3}\times\mathcal{E}(Z_*^2) & Z_*^{-2}\times\mathcal{E}(Z_*^2)\\
Z_*^{-2}\times\mathcal{E}(Z_*^2) & Z_*^{-1}\times\mathcal{E}(Z_*^2)
\end{pmatrix}.
\label{eq:sigmaZ_constDz}
\end{equation}
This replaces Eq.~\eqref{sigmaZ} sufficiently close to the CEP whenever \(D_z\neq0\). By contrast, when the \(Z_*^2\) part of \((D_{\perp})_{22}\) dominates, the original scaling \eqref{sigmaZ} is recovered and the effect of \(D_z\) is only subleading.

Physically, a finite \(D_z\) injects noise directly into the CEP-sensitive channel. Thus, when the coalescing direction is \({\bf e}_z\), dephasing in the transverse \((x,y)\) directions is suppressed, whereas dephasing with a \(z\) component remains relevant and cuts off the CEP-induced squeezing.

\newpage
\newcommand{\ii}{\mathrm{i}}
\newcommand{\PT}{\mathcal{P}\mathcal{T}}
\newcommand{\Diss}[2]{\left(#1\,#2\,#1^\dagger-\frac12\{#1^\dagger #1,#2\}\right)}
\newcommand{\Cov}{\mathrm{Cov}}

\newpage
\AppSection{L-$\mathcal{PT}$ symmetric collective-spin model}

We now apply the general analysis to the collective-spin model used in the main text,
\begin{equation}
H_S
=
S\left(g m_x+\frac{\omega}{2}m_z^2\right),
\qquad
L_-=\sqrt{\kappa S}\,m_-,
\qquad
m_-:=m_x-i m_y .
\label{eq:example_model_supp}
\end{equation}
Equivalently, \(H_S=gS_x+\omega S_z^2/(2S)\) and
\(L_-=\sqrt{\kappa/S}\,S_-\).
This model exhibits an L-$\mathcal{PT}$ phase transition with a CEP~\cite{Nakanishi3}.

For \(\bm M=(X,Y,Z)^T=\langle \bm m\rangle\), the mean-field equation is
\begin{equation}
\dot{\bm M}=\bm g(\bm M),
\qquad
\bm g(\bm M)=
\begin{pmatrix}
\kappa ZX-\omega ZY\\
-gZ+\kappa ZY+\omega ZX\\
gY-\kappa(1-Z^2)
\end{pmatrix}.
\label{eq:MF_example}
\end{equation}
In the PT-broken phase, \(Z_*\neq0\), the fixed point is
\begin{equation}
X_*=\frac{g\omega}{\kappa^2+\omega^2},
\qquad
Y_*=\frac{g\kappa}{\kappa^2+\omega^2},
\qquad
Z_*=\pm \sqrt{1-\frac{g^2}{\kappa^2+\omega^2}} .
\label{eq:PTbroken_fixed_point_example}
\end{equation}
Thus the PT-broken solution exists for
\begin{equation}
g^2<\kappa^2+\omega^2,
\end{equation}
and the transition is reached at
\begin{equation}
g=g_c:=\sqrt{\kappa^2+\omega^2},
\qquad
Z_*\to0.
\label{eq:gc_example}
\end{equation}
Below we take the stable branch \(Z_*<0\).

\AppSubsection{Linearized dynamics}

From Eq.~\eqref{eq:MF_example}, the full Jacobian is
\begin{equation}
J(\bm M)=
\begin{pmatrix}
\kappa Z & -\omega Z & \kappa X-\omega Y\\
\omega Z & \kappa Z & -g+\kappa Y+\omega X\\
0 & g & 2\kappa Z
\end{pmatrix}.
\label{eq:J_full_example}
\end{equation}
At the PT-broken fixed point, Eq.~\eqref{eq:PTbroken_fixed_point_example} implies
\begin{equation}
\kappa X_*-\omega Y_*=0,
\qquad
-g+\kappa Y_*+\omega X_*=0.
\end{equation}
Hence the Jacobian reduces to
\begin{equation}
J_*=
\begin{pmatrix}
\kappa Z_* & -\omega Z_* & 0\\
\omega Z_* & \kappa Z_* & 0\\
0 & g & 2\kappa Z_*
\end{pmatrix}.
\label{eq:J_star_example}
\end{equation}

After rotating to the frame in which the mean spin points along the \(z'\) axis, the transverse block becomes
\begin{equation}
J'_\perp(Z_*)=
\begin{pmatrix}
\kappa Z_* & -\omega\\
\omega Z_*^2 & \kappa Z_*
\end{pmatrix}.
\label{eq:Jperp_example}
\end{equation}
For any finite \(\omega\neq0\), this matrix approaches
\begin{equation}
J'_\perp(Z_*\to0)
=
\begin{pmatrix}
0 & -\omega\\
0 & 0
\end{pmatrix},
\end{equation}
which is a Jordan block. Thus the CEP appears directly in the tangent-plane dynamics for finite \(\omega\).

\AppSubsection{Diffusion matrix and covariance}

For the collective decay channel \(L_-\), the transverse diffusion matrix in the same rotated frame is
\begin{equation}
D'_\perp(Z_*)=
\begin{pmatrix}
2\kappa & 0\\
0 & 2\kappa Z_*^2
\end{pmatrix}.
\label{eq:Dperp_example}
\end{equation}
In particular, there is no \(Z_*^0\) diffusion component in the second transverse channel, namely \(D_z=0\).

Solving the reduced Lyapunov equation
\begin{equation}
J'_\perp\Sigma_\perp+\Sigma_\perp J_\perp^{\prime T}+D'_\perp=0
\label{eq:Lyapunov_example}
\end{equation}
on the stable branch \(Z_*<0\), one obtains
\begin{equation}
\Sigma_\perp^\prime(Z_*)=
\begin{pmatrix}
-\dfrac{1}{Z_*} & 0\\[2mm]
0 & -Z_*
\end{pmatrix}.
\label{eq:Sigma_perp_example}
\end{equation}
Therefore,
\begin{equation}
\lambda_{\max}(\Sigma_\perp^\prime)\sim |Z_*|^{-1},
\qquad
\lambda_{\min}(\Sigma_\perp^\prime)\sim |Z_*|.
\label{eq:lambda_scaling_example}
\end{equation}
The CEP-enhanced squeezing scaling thus follows explicitly in this model.

Importantly, Eq.~\eqref{eq:Sigma_perp_example} does not depend on \(\omega\). Hence the same scaling survives not only in the limit \(\omega\to0\), but also at \(\omega=0\) exactly. What changes at \(\omega=0\) is not the scaling law itself, but the geometric origin of the tangent-plane description: the projected \(2\times2\) dynamics is no longer a Jordan block, even though the full \(3\times3\) Jacobian remains defective.

\AppSubsection{Effect of additional dephasing}

We now add collective dephasing channels
\begin{equation}
L_\alpha=\sqrt{\gamma_\alpha S}\,m_\alpha,
\qquad
\alpha=x,y,z .
\end{equation}
These Hermitian linear jump operators do not modify the mean-field equation, and hence the fixed point
\((X_*,Y_*,Z_*)\) and the CEP condition remain unchanged. Their effect appears only in the diffusion matrix.

In the rotated frame used in Eq.~\eqref{eq:Jperp_example}, the transverse diffusion matrix can be written as
\begin{equation}
D'_\perp(Z_*)=
\begin{pmatrix}
d_{11} & d_{12}Z_*\\
d_{12}Z_* & \gamma_z+d_{22}Z^2_*
\end{pmatrix},
\label{eq:Dperp_all_dephasing}
\end{equation}
with
\begin{align}
d_{11}
&=
2\kappa
+
\frac{\gamma_x\kappa^2+\gamma_y\omega^2}{\kappa^2+\omega^2},
\label{eq:d11_all_dephasing}
\\
d_{12}
&=
\frac{\kappa\omega}{\kappa^2+\omega^2}
(\gamma_x-\gamma_y),
\label{eq:d12_all_dephasing}
\\
d_{22}
&=
\left[
2\kappa
+
\frac{\gamma_x\omega^2+\gamma_y\kappa^2}{\kappa^2+\omega^2}-\gamma_z
\right].
\label{eq:d22_all_dephasing}
\end{align}
Thus the \(x\)- and \(y\)-dephasing channels only contribute to the analytic \(Z_*\)-dependent part of \((D'_\perp)_{22}\), whereas the \(z\)-dephasing channel produces a finite \(Z_*^0\) contribution in \((D'_\perp)_{22}\). In the notation of the general discussion, this constant contribution is
\begin{equation}
D_z=\gamma_z .
\end{equation}

Solving the Lyapunov equation gives
\begin{align}
(\Sigma_\perp^\prime)_{11}
&=
-
\frac{
\left[(2\kappa^2+\omega^2)d_{11}
+2\kappa\omega d_{12}
+\omega^2 d_{22}\right]Z_*^2
+\gamma_z\omega^2
}
{
4\kappa(\kappa^2+\omega^2)Z_*^3
},
\\
(\Sigma_\perp^\prime)_{12}
&=
\frac{
\omega d_{11}Z_*^2
-2\kappa d_{12}Z_*^2
-\omega d_{22}Z_*^2
-\gamma_z\omega
}
{
4(\kappa^2+\omega^2)Z_*^2
},
\\
(\Sigma_\perp^\prime)_{22}
&=
-
\frac{
\left[\omega^2 d_{11}
-2\kappa\omega d_{12}
+(2\kappa^2+\omega^2)d_{22}\right]Z_*^2
+(2\kappa^2+\omega^2)\gamma_z
}
{
4\kappa(\kappa^2+\omega^2)Z_*
}.
\end{align}

The scaling follows directly from above equations. If \(\gamma_z=0\), then \(d_{22}=O(Z_*^2)\) and \(d_{12}=O(Z_*)\), so the original CEP scaling is preserved:
\begin{equation}
\Sigma_\perp^\prime
\sim
\begin{pmatrix}
|Z_*|^{-1} & O(1)\\
O(1) & |Z_*|
\end{pmatrix}.
\end{equation}
Thus \(x\)- and \(y\)-dephasing do not change the leading powers of the squeezed and anti-squeezed variances.

By contrast, when \(\gamma_z\neq0\), \(d_{22}\) contains the constant contribution \(D_z=\gamma_z\). For finite \(\omega\neq0\), this term dominates sufficiently close to the CEP and gives
\begin{equation}
\Sigma_\perp^\prime
\sim
\gamma_z
\begin{pmatrix}
|Z_*|^{-3} & |Z_*|^{-2}\\
|Z_*|^{-2} & |Z_*|^{-1}
\end{pmatrix}.
\label{eq:Sigma_Dz_scaling_concrete}
\end{equation}

\AppSubsection{Finite-size scaling}

We estimate the finite-size rounding of the CEP scaling from a Ginzburg-type self-consistency criterion~\cite{Rikvold1993Critical,ColonnaRomano2014Anomalous}. 
Throughout this subsection, \(Z_*\) denotes the thermodynamic mean-field order parameter on a selected PT-broken branch. 
For finite \(S\), we define \(Z_S\) as the corresponding steady-state order parameter, with the sign chosen such that \(Z_S>0\).

For \(D_z=0\), the transverse covariance obeys
\begin{align}
\lambda_{\max}(\Sigma'_\perp)\sim |Z_*|^{-1},
\qquad
\lambda_{\min}(\Sigma'_\perp)\sim |Z_*|.
\label{eq:fss_lambda_scaling_short}
\end{align}
The physical fluctuation amplitude along the anti-squeezed direction is therefore
\begin{align}
\Delta x_{\mathrm{asq}}
\sim
\sqrt{\frac{\lambda_{\max}(\Sigma'_\perp)}{S}}
\sim
\frac{1}{\sqrt{S|Z_*|}} .
\label{eq:asq_width_short}
\end{align}
The single-branch Gaussian theory is self-consistent only when this fluctuation remains smaller than the distance \(|Z_*|\) from the CEP. 
Since the anti-squeezed direction approaches the coalescing direction as \(Z_*\to0\), this gives
\begin{align}
\Delta x_{\mathrm{asq}}\ll |Z_*|
\qquad\Longrightarrow\qquad
S|Z_*|^3\gg1 .
\label{eq:ginzburg_short}
\end{align}
Thus the finite-size crossover is determined by
\begin{align}
S|Z_S|^3\sim1,
\qquad
|Z_S|\sim S^{-1/3}.
\label{eq:ZG_short}
\end{align}

\begin{figure*}[t]
\centering
\includegraphics[width=0.8\linewidth]{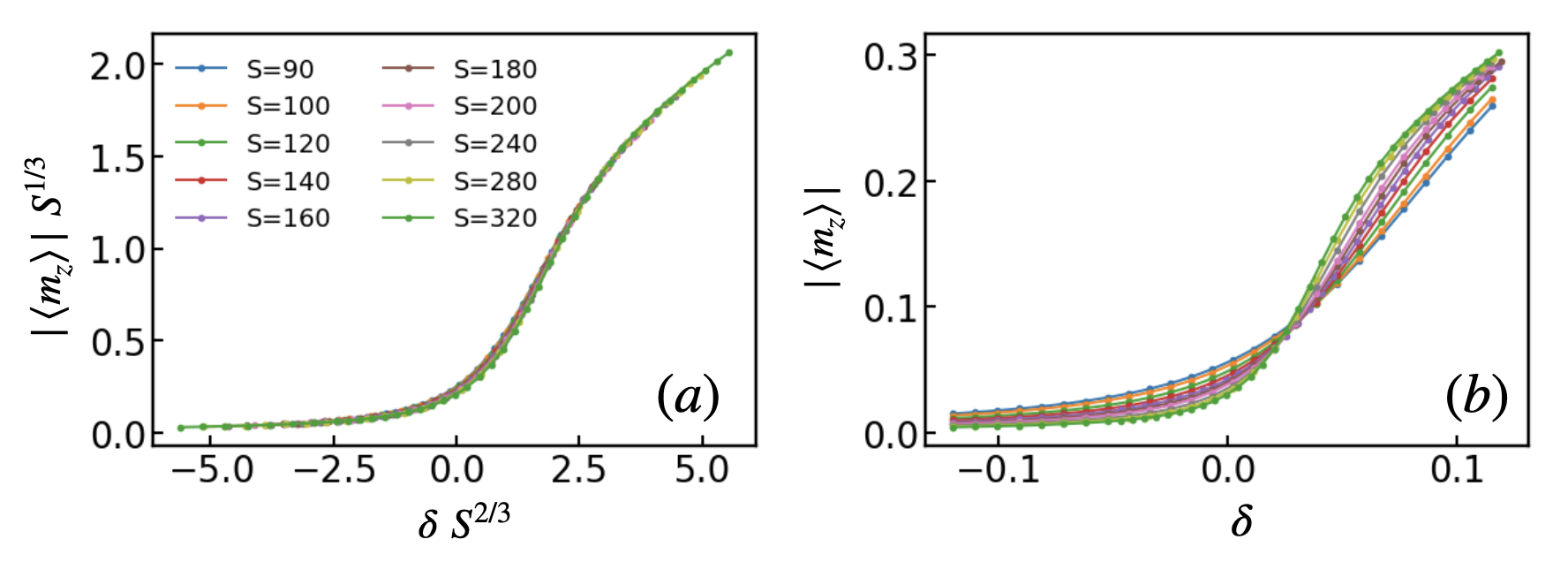}
\caption{
Finite-size scaling of the order parameter near the CEP.
(a) Scaling collapse of \(|\braket{m_z}| S^{1/3}\) plotted against \(\delta S^{2/3}\).
(b) Unscaled finite-\(S\) order parameter \(|\braket{m_z}|\) as a function of \(\delta\).
The collapse supports
\(Z_S\delta=S^{-1/3}\mathcal Z(S^{2/3}\delta)\).
}
\label{fig:Z_fss}
\end{figure*}

From Eq.~\eqref{eq:gc_example}, we obtain \(|Z_*|\sim\delta^{1/2}\) with $\delta:=(\kappa-\kappa_c)/\omega$
Combining this with Eq.~\eqref{eq:ZG_short}, we obtain
\begin{align}
\delta_S\sim S^{-2/3}.
\end{align}
The corresponding finite-size scaling form of the order parameter is
\begin{align}
Z_S(\delta)
=
S^{-1/3}{\cal Z}(S^{2/3}\delta),
\qquad
{\cal Z}(u)\sim u^{1/2}\quad (u\gg1).
\label{eq:ZS_scaling_short}
\end{align}
This is consistent with the scaling exponents for $\omega=0$ case in Ref.~\cite{Montenegro}. 
Figure~\ref{fig:Z_fss} verifies this scaling numerically: the unscaled curves sharpen near the CEP, while \(Z S^{1/3}\) collapses when plotted against \(\delta S^{2/3}\).

Since the thermodynamic Gaussian theory gives
\(\xi_S^2=\lambda_{\min}(\Sigma'_\perp)\sim |Z_*|\), the same Ginzburg scale fixes the leading algebraic powers of the squeezing cutoff:
\begin{align}
\xi_S^2(\delta;S)
=
S^{-1/3}\Phi_0(S^{2/3}\delta),
\qquad
\Phi_0(u)\sim u^{1/2}\quad (u\gg1).
\label{eq:xi_fss_short}
\end{align}
This gives the algebraic cutoff
\begin{align}
(\xi_S^2)_{\mathrm{cutoff}}\sim S^{-1/3}.
\end{align}

This algebraic Ginzburg estimate does not fix possible marginal logarithmic corrections. 
Following the main text, we introduce the logarithmically corrected effective size
\begin{align}
S_{\rm eff}:=\frac{S}{\log S},
\end{align}
motivated by the logarithmic optimization known for $\omega=0$ case~\cite{BarberenaRey2024}. 
For the inverse squeezing
\begin{align}
G:=\frac{1}{\xi_S^2},
\end{align}
we use
\begin{align}
G(\delta;S)
=
S_{\rm eff}^{1/3}
{\cal G}\!\left(\delta S_{\rm eff}^{2/3}\right).
\label{eq:G_log_fss_short}
\end{align}
This is the scaling form tested in Fig.~\ref{fig:squeezing_CEP}(f) in the main text, where the data are plotted as
\begin{align}
G(S/\log S)^{-1/3}
\quad\text{against}\quad
\delta(S/\log S)^{2/3}.
\end{align}
The collapse supports the logarithmically corrected finite-size scaling of the inverse squeezing.

\newpage

\AppSection{Example: spin-boson model}

In this section, we show that the CEP-controlled squeezing mechanism also appears in a spin-boson realization with a lossy cavity mode.

We consider
\begin{equation}
\dot\rho
=
-\ii[H,\rho]
+
\kappa\,\Diss{a}\rho,
\qquad
H
=
gS_x
+
\frac{\lambda}{\sqrt N}\left(a^\dagger S_-+aS_+\right)
+
\frac{\omega}{N}S_z^2 .
\label{eq:SB_ME}
\end{equation}
Here \(g\) is the transverse field, \(\omega\) is the nonlinear spin interaction, \(\lambda\) is the spin-boson coupling, and \(\kappa\) is the cavity-loss rate.

Following Sec.~A, we denote the spin-boson variables by
\begin{equation}
\bm{x}:=(x,y,z,q,p)^T,
\qquad
\bm{X}:=(X,Y,Z,Q,P)^T .
\end{equation}
The mean-field equations are
\begin{align}
\dot X
&=
\sqrt2\,\lambda PZ-\omega ZY,
\label{eq:SB_MFx}
\\
\dot Y
&=
-gZ-\sqrt2\,\lambda QZ+\omega ZX,
\label{eq:SB_MFy}
\\
\dot Z
&=
gY+\sqrt2\,\lambda QY-\sqrt2\,\lambda PX,
\label{eq:SB_MFz}
\\
\dot Q
&=
-\frac{\lambda}{\sqrt2}Y-\frac{\kappa}{2}Q,
\qquad
\dot P
=
\frac{\lambda}{\sqrt2}X-\frac{\kappa}{2}P.
\label{eq:SB_MFqp}
\end{align}
Solving the fixed-point equations together with
\(X_*^2+Y_*^2+Z_*^2=1\), we obtain the symmetry-broken
fixed point
\begin{equation}
X_*=\frac{g\omega\kappa^2}{\Delta},
\qquad
Y_*=\frac{2g\kappa\lambda^2}{\Delta},
\qquad
Z_*=\pm\sqrt{1-\frac{g^2\kappa^2}{\Delta}},
\label{eq:SB_fixed_point}
\end{equation}
with
\begin{equation}
Q_*=-\frac{\sqrt2\,\lambda}{\kappa}Y_*,
\qquad
P_*=\frac{\sqrt2\,\lambda}{\kappa}X_*,
\qquad
\Delta:=4\lambda^4+\omega^2\kappa^2 .
\end{equation}
This branch exists for \(g^2\kappa^2<\Delta\), and the
continuous transition is reached at
\begin{equation}
g_c(\omega)=\frac{\sqrt{4\lambda^4+\omega^2\kappa^2}}{\kappa}.
\label{eq:SB_gc}
\end{equation}
Equivalently, if \(\omega\) is used as the control parameter at fixed \(g\),
\begin{equation}
\omega_c=\sqrt{g^2-\frac{4\lambda^4}{\kappa^2}},
\label{eq:SB_omegac}
\end{equation}
which is real for \(g^2>4\lambda^4/\kappa^2\).

The full lab-frame Jacobian at the fixed point is
\begin{equation}
J_*
=
\begin{pmatrix}
0 & -\omega Z_* & 0 & 0 & \sqrt2\,\lambda Z_*\\[4pt]
\omega Z_* & 0 & 0 & -\sqrt2\,\lambda Z_* & 0\\[8pt]
-\dfrac{2g\omega\kappa\lambda^2}{\Delta}
&
\dfrac{g\omega^2\kappa^2}{\Delta}
&
0
&
\dfrac{2\sqrt2\,g\kappa\lambda^3}{\Delta}
&
-\dfrac{\sqrt2\,g\omega\kappa^2\lambda}{\Delta}
\\[10pt]
0 & -\dfrac{\lambda}{\sqrt2} & 0 & -\dfrac{\kappa}{2} & 0\\[8pt]
\dfrac{\lambda}{\sqrt2} & 0 & 0 & 0 & -\dfrac{\kappa}{2}
\end{pmatrix}.
\label{eq:SB_Jacobian5x5}
\end{equation}
At the transition point \(Z_*=0\),
\begin{equation}
\left.
\det(s\mathbbm{1}-J_*)
\right|_{Z_*=0}
=
s^3\left(s+\frac{\kappa}{2}\right)^2 .
\label{eq:SB_charpoly}
\end{equation}
For \(\lambda\neq0\), the zero-eigenvalue eigenspace is two-dimensional.
Hence the zero-eigenvalue sector contains a nontrivial \(J_2(0)\) block,
and the continuous transition is a CEP.

For the scaled fluctuations
\begin{equation}
\eta_i=\sqrt N\,(x_i-X_{i,*}),
\qquad
\bm{\xi}:=(\eta_x,\eta_y,\eta_z,\eta_q,\eta_p)^T,
\end{equation}
the lab-frame diffusion matrix is
\begin{equation}
D_{\rm SB}
=
\frac{\kappa}{2}\,\mathrm{diag}(0,0,0,1,1).
\label{eq:SB_Diffusion5x5}
\end{equation}

We now compute the spin covariance in the physical tangent-cavity sector.
Applying the aligned-frame projection introduced in Sec.~B.4, we remove
the radial spin fluctuation and keep
\begin{equation}
\bm{\eta}'_{\rm SB}:=(\eta_x',\eta_y',\eta_q',\eta_p')^T,
\qquad
\eta_q'=\eta_q,\quad \eta_p'=\eta_p .
\end{equation}
The linearized dynamics is
\begin{equation}
\dot{\bm{\eta}}'_{\rm SB}
=
J'_{\rm SB}\bm{\eta}'_{\rm SB}
+
\bm{\zeta}',
\end{equation}
with
\begin{equation}
J'_{\rm SB}
=
\begin{pmatrix}
0
&
-\omega
&
-\sqrt2\,\lambda\dfrac{Y_*}{r_{\perp,*}}
&
\sqrt2\,\lambda\dfrac{X_*}{r_{\perp,*}}
\\[8pt]
\omega Z_*^2
&
0
&
-\sqrt2\,\lambda\dfrac{X_*Z_*}{r_{\perp,*}}
&
-\sqrt2\,\lambda\dfrac{Y_*Z_*}{r_{\perp,*}}
\\[8pt]
-\dfrac{\lambda}{\sqrt2}\dfrac{Y_*Z_*}{r_{\perp,*}}
&
-\dfrac{\lambda}{\sqrt2}\dfrac{X_*}{r_{\perp,*}}
&
-\dfrac{\kappa}{2}
&
0
\\[8pt]
\dfrac{\lambda}{\sqrt2}\dfrac{X_*Z_*}{r_{\perp,*}}
&
-\dfrac{\lambda}{\sqrt2}\dfrac{Y_*}{r_{\perp,*}}
&
0
&
-\dfrac{\kappa}{2}
\end{pmatrix},
\qquad
r_{\perp,*}:=\sqrt{X_*^2+Y_*^2}.
\label{eq:SB_tangent_drift}
\end{equation}
The corresponding diffusion matrix is
\begin{equation}
D'_{\rm SB}
=
\frac{\kappa}{2}\,
\mathrm{diag}(0,0,1,1).
\label{eq:SB_tangent_diffusion}
\end{equation}
The radial zero mode has been removed. For finite \(\omega\neq0\), the
tangent-cavity sector remains defective:
\begin{equation}
\left.
\det(s\mathbbm{1}-J'_{\rm SB})
\right|_{Z_*=0}
=
s^2\left(s+\frac{\kappa}{2}\right)^2,
\qquad
\left.
\mathrm{Ker}\,J'_{\rm SB}
\right|_{Z_*=0}
=
\mathrm{span}\{(1,0,0,0)^T\}.
\label{eq:SB_tangent_charpoly}
\end{equation}
Thus the zero-eigenvalue sector contains a nontrivial \(J_2(0)\) block,
whose eigenvector is the \(\xi_x'\) direction.

The covariance matrix \(\Sigma'_{\rm SB}\) obeys
\begin{equation}
\dot{\Sigma}'_{\rm SB}
=
J'_{\rm SB}\Sigma'_{\rm SB}
+
\Sigma'_{\rm SB}(J'_{\rm SB})^T
+
D'_{\rm SB}.
\label{eq:SB_Lyapunov_tangent}
\end{equation}
On the stable symmetry-broken branch \(Z_*<0\), the stationary solution is
\begin{equation}
\Sigma'_{{\rm SB},*}
=
\mathrm{diag}
\left(
-\frac{1}{Z_*},
-Z_*,
\frac{1}{2},
\frac{1}{2}
\right),
\qquad
J'_{\rm SB}\Sigma'_{{\rm SB},*}
+
\Sigma'_{{\rm SB},*}(J'_{\rm SB})^T
+
D'_{\rm SB}=0 .
\label{eq:SB_covariance_solution}
\end{equation}
Therefore the transverse spin covariance is
\begin{equation}
\Sigma'_{\perp,*}
=
\begin{pmatrix}
-Z_*^{-1} & 0\\
0 & -Z_*
\end{pmatrix}.
\label{eq:SB_spin_covariance}
\end{equation}
For finite \(\omega\neq0\), the anti-squeezed direction is the defective
\(\eta_x'\) direction, while the squeezed direction is \(\eta_y'\). Hence
\begin{equation}
\xi_S^2
=
\lambda_{\min}(\Sigma'_{\perp,*})
=
-Z_*
=
\sqrt{
1-
\frac{g^2\kappa^2}
{4\lambda^4+\omega^2\kappa^2}
}.
\label{eq:SB_xi_result}
\end{equation}
Thus the CEP yields
\begin{equation}
\xi_S^2\to0,
\quad
Z_*\to0 .
\label{eq:SB_CEP_squeezing_limit}
\end{equation}
\end{document}